\documentclass[twocolumn,10pt]{asme2e}

\conffullname{the 41st International Conference on Ocean, Offshore and Arctic Engineering}
\confshortname{OMAE 2022}
\confdate{5-10}
\confmonth{June}
\confyear{2022}
\confcity{Hamburg}
\confcountry{Germany}
\papernum{OMAE2022-81006}

\title{Control Co-design of a Hydrokinetic Turbine with Open-loop Optimal Control}

\author{Boxi Jiang 
    \affiliation{
	Naval Architecture and Marine Eng.\\
	University of Michigan\\
	Ann Arbor, Michigan 48109\\
    Email: boxij@umich.edu
    }	
}
\author{Mohammad Reza Amini\thanks{Address all correspondence to this author.}
    \affiliation{
	Naval Architecture and Marine Eng.\\
	University of Michigan\\
	Ann Arbor, Michigan 48109\\
    Email: mamini@umich.edu
    }	
}

\author{Yingqian Liao
    \affiliation{
	Aerospace Engineering\\
	University of Michigan\\
	Ann Arbor, Michigan 48109\\
    Email: yqliao@umich.edu
    }}

\author{Joaquim R.R.A. Martins
    \affiliation{
	Aerospace Engineering\\
	University of Michigan\\
	Ann Arbor, Michigan 48109\\
    Email: jrram@umich.edu
    }}

\author{Jing Sun
    \affiliation{
	Naval Architecture and Marine Eng.\\
	University of Michigan\\
	Ann Arbor, Michigan 48109\\
    Email: jingsun@umich.edu
    }}
    
\usepackage{graphicx}
\usepackage{amsmath}
\usepackage{array}
\usepackage{amssymb}
\usepackage{verbatim}
\usepackage{cleveref}
\usepackage{color}
\usepackage{xcolor}
\usepackage{relsize}
\usepackage{tikz}
\usepackage{fancyhdr}
\usepackage{graphics}
\usepackage{setspace}
\usepackage{epstopdf}
\usepackage{multirow}
\usepackage{lipsum}
\usepackage{textcomp}
\usepackage{mathtools}
\usepackage{cite}
\usepackage{setspace}
\usepackage{booktabs}
\usepackage[noend]{algpseudocode}
\usepackage{flushend}
\usepackage{multicol}
\usepackage{float}
\usepackage{mathtools,bm}
\usepackage[noend]{algpseudocode}
\usepackage{booktabs}
\usepackage{dblfloatfix}
\usepackage{mathrsfs}
\usepackage{mathtools}
\usepackage{optidef}
\usepackage[ruled,vlined]{algorithm2e}

\begin{document}

\maketitle    
\begin{abstract}
{\it This paper introduces a control co-design (CCD) framework to simultaneously explore the physical parameters and control spaces for a hydro-kinetic turbine (HKT) rotor optimization. The optimization formulation incorporates a coupled dynamic-hydrodynamic model to maximize the rotor power efficiency for various time-variant flow profiles. The open-loop optimal control is applied for maximum power tracking, and the blade element momentum theory (BEMT) is used to model the hydrodynamics. Case studies with different control constraints are investigated for CCD. 
Sensitivity analyses were conducted with respect to different flow profiles and initial geometries.
Comparisons are made between CCD and the sequential process, with physical design followed by a control design process under the same conditions. The results demonstrate the benefits of CCD and reveal that, with control constraints, CCD leads to increased energy production compared to the design obtained from the sequential design process.}
\end{abstract}

\begin{nomenclature}
\entry{\(c\)}{\ Chord length, m}
\entry{$\alpha$}{Twist angle, degree}
\entry{\(v\)}{\ Flow velocity, m/s}
\entry{$\omega$}{Turbine rotating speed, rad/s}
\entry{\(r\)}{\ Turbine radius, m}
\entry{$\lambda$}{Tip speed ratio; $\frac{\omega r}{v}$}
\entry{\(u\)}{\ Control load, Nm}
\entry{\(Q\)}{\ Fluid-induced torque, Nm}
\entry{$C_p$}{Power coefficient}
\end{nomenclature}

\section{INTRODUCTION}
The design of a hydro-kinetic turbine (HKT) is often performed sequentially, with the physical design space (such as the blade geometry) being first explored with a focus on steady-state performance~\cite{leishman2002challenges}, followed by the control design addressing the transient responses once the physical parameters are defined. For highly dynamic and interactive systems such as HKT, the sequential design approach cannot fully explore the synergy between the physical entity and its control system dynamics, leading to sub-optimal design or overly constrained control solutions~\cite{martins2021engineering}. On the other hand, control co-design (CCD) can simultaneously optimize physical design variables and control parameters for best performance and trade-off management~\cite{garcia2019control}.

Recent research on CCD for renewable energy generation systems has shown promising improvements in power production, load mitigation, and cost reduction. As for floating offshore wind turbines (FOWT), Deshmukh and Allison~\cite{deshmukh2013simultaneous} formulated a simultaneous structural and control system design and presented significant performance improvements over sequential design approaches through case studies. Du and Bilgen~\cite{du2020control} investigated the coupling effect between the physical space, defined by the pre-cone angle, and the pitch control to reduce the blade root bending moment through CCD. Pao et al.~\cite{pao2021control} developed an aero-structural-control co-design process and demonstrated a 25\% reduction in levelized cost of energy (LCOE) through a case study of a 13MW FOWT. Sundarrajan et al.~\cite{sundarrajan2021open} applied a linear parameter-varying model with CCD and achieved a mass reduction in the wind turbine plant while satisfying power demands and maintaining stability constraints. As for the hydrokinetic energy devices, Naik and Vermillion~\cite{naik2021fused} developed a CCD framework for optimizing an energy-harvesting ocean kite subject to structural constraints. Coe et al.~\cite{coe2020initial} conducted a conceptual demonstration of control co-design for wave energy converter (WEC). As for the HKT, the application of CCD is still at an early age, with no control co-designed systems reported in the open literature to the best knowledge of the authors.

This paper introduces a CCD framework for the rotor design optimization of an HKT. A coupled dynamic-hydrodynamic model is introduced, based on which the CCD problem is formulated to maximize the energy production for different time-variant flow profiles. The CCD-based HKT design optimization is investigated within the below-rated speed operating zone, and the objective is to maximize the HKT power production. In this case study, we consider twist angle and chord length along the blade as the physical design parameters. CCBlade~\cite{ning2013ccblade}, a blade element momentum theory-based package, is used to model the hydrodynamics. Open-loop optimal torque control is applied for maximum HKT power production. 
The control variable and other time-varying trajectories are simulated and optimized using open-source software Dymos~\cite{falck2021dymos}. CCD under different control constraints is investigated together with sensitivity analysis to different flow profiles and initial geometries. 
Details on physical designs, including power production, trajectories of the resulting optimal control, and turbine rotating speed, are reported and discussed. Geometries and power performance of CCD designs are compared against those of sequential design under the same conditions including flow profiles and control constraints. The results show that in the presence of control constraints, CCD leads to increased energy production, compared to the sequential (design then control) methodology.

The contributions of this study are tri-fold: First, a CCD framework is developed for an HKT with open-loop optimal control. Second, a comprehensive performance evaluation of the CCD results, compared with the sequential design, is carried out to reveal interactions between the two design spaces and provide insights for integration. Finally, the sensitivity of the CCD results to the flow profile and initial HKT geometry is analyzed.

\vspace{-6pt}
\section{ROTOR BLADE DESIGN AND CONTROL FOR MAXIMUM POWER PRODUCTION}
\begin{figure}[h!]
      \includegraphics[scale=0.9]{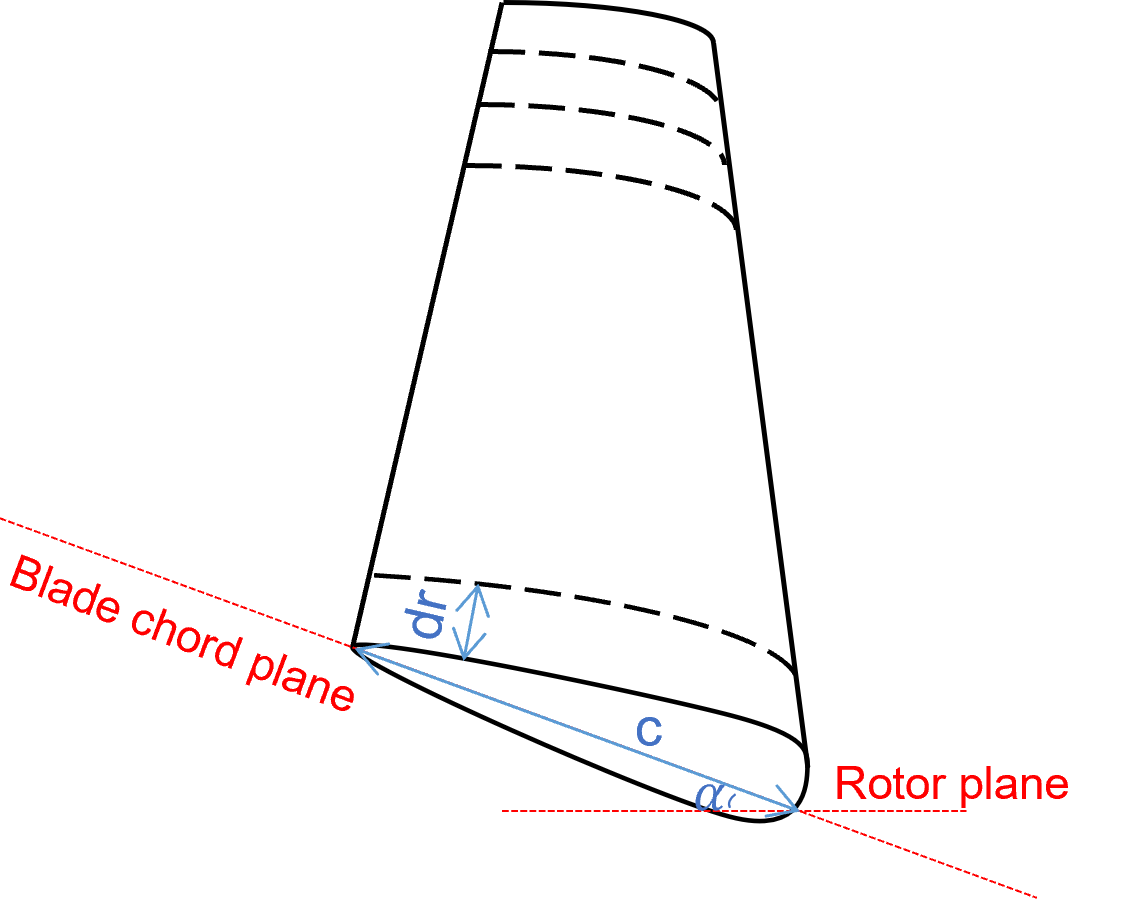}\vspace{-6pt}
      \caption{Illustration of blade geometry and model in BEM.}\vspace{-10pt}
            \label{figure: geom}
\end{figure}

The design objective is to maximize the HKT power production in a time-varying flow condition through the selection of the physical design of the rotor blades and the control. The baseline turbine is a three-bladed horizontal-axis turbine. A quasi-static model based on Blade Element Momentum (BEM)~\cite{ning2014simple} theory is used to predict rotor performances with specific geometric design parameters. BEM divides a blade into several sections and each section is referred to as a blade element, as shown in Figure~\ref{figure: geom}. Detailed problem formulation, including the design operating condition, the design parameters, the open-loop optimal control, and the baseline model will be introduced.
\vspace{-10pt}

\subsection{Baseline Turbine Model}
The baseline geometry of this study is adopted from the NREL 5MW wind turbine~\cite{jonkman2009definition}. A geometric scaling ratio of 1:10 is applied so that the scaled model has a rated power of 100 kW, which better fits the power requirements for the grid-connection application.  The advance ratio is maintained to be the same between the NREL 5MW model and the scaled model. Since the scaled model is an HKT, the mass is further scaled based on the power to mass ratio considering the structural strength~\cite{griffith2017gravo}. Also, the rotor performance is assumed
to be Reynolds invariant. The detailed parameters of the NREL 5MW model and the scaled HKT model are listed in Table~\ref{table_1}.

\begin{table}[h!]
\caption{Specifications of the NREL 5MW turbine and the scaled HKT model}
\label{table_1}
\begin{center}
\begin{tabular}{l c c}
\hline
 & \textbf{NREL 5MW} & \textbf{Scaled}\\
 & \textbf{Turbine}\cite{jonkman2009definition} & \textbf{HKT Model} \\
\hline
Rated Power ($kW$) & 5000  & 100\\
Rated Speed ($m/s$) & 11.4 & 1.6\\
Rotor diameter ($m$) &126 &12.6\\
Hub diameter($m$) &3 & 0.3\\
Number of blades (-) &3 &3\\
Rotor mass ($kg$) &17740 &354.8 \\
Rotor inertia ($kgm^{2}$) &11776047 &2234\\
\hline
\end{tabular}\vspace{-10pt}
\end{center}
\end{table}

\vspace{-6pt}
\subsection{Design Parameters}
The HKT geometry variables explored in this study are the twist angles and chord lengths.
Other parameters, such as the thickness to chord length ratio along the radius and blade radius, are kept constant. The baseline foil is the DU21A17 profile used in the NREL 5MW turbine~\cite{jonkman2009definition}. The design parameters of HKT rotors are defined by discretizing the blades into 10 different segments. The three segments near the root have cylindrical geometries, whose radius is fixed during design for maintaining structural safety. 

\vspace{-6pt}
\subsection{Open-loop Optimal Control}
When the HKT is operating in the below-rated region, a variable-speed control is generally applied to track the turbine maximum power coefficient locus \cite{kim2012maximum}, which is defined by the physical design. In this study, an open-loop optimal torque control~\cite{kim2010fast} is employed, assuming that the generator torque can be controlled directly for optimal power production and the inflow velocity profile is known with sensors. The continuous optimal control load trajectory is discretized in time using third-order Legendre-Gauss-Lobatto collocation method and then solved numerically as a nonlinear programming problem in Dymos\cite{falck2021dymos}. Generator torque at each time instant is treated as an optimization variable. No feedback is used. Different scenarios where the generator torque is subject to physical constraints are further analyzed in Section 4.

\vspace{-6pt}
\subsection{Optimizer}
We used SNOPT~\cite{gill2005snopt}, a well-known sequential quadratic programming algorithm, as the optimizer with both feasibility tolerance and optimality tolerance of $10^{-6}$ for design optimization, optimal control, and CCD.

\vspace{-6pt}
\section{DESIGN OPTIMIZATION APPROACHES}
Since we are interested in understanding how CCD brings additional benefits to the system design. We perform optimizations using two approaches and make comparisons between the results. The first approach is the conventional sequential design and optimization approach. The second one is CCD.

\vspace{-6pt}
\subsection{Sequential Design and Optimization Formulation} 
\begin{figure}[h!]
      \includegraphics[scale=0.5]{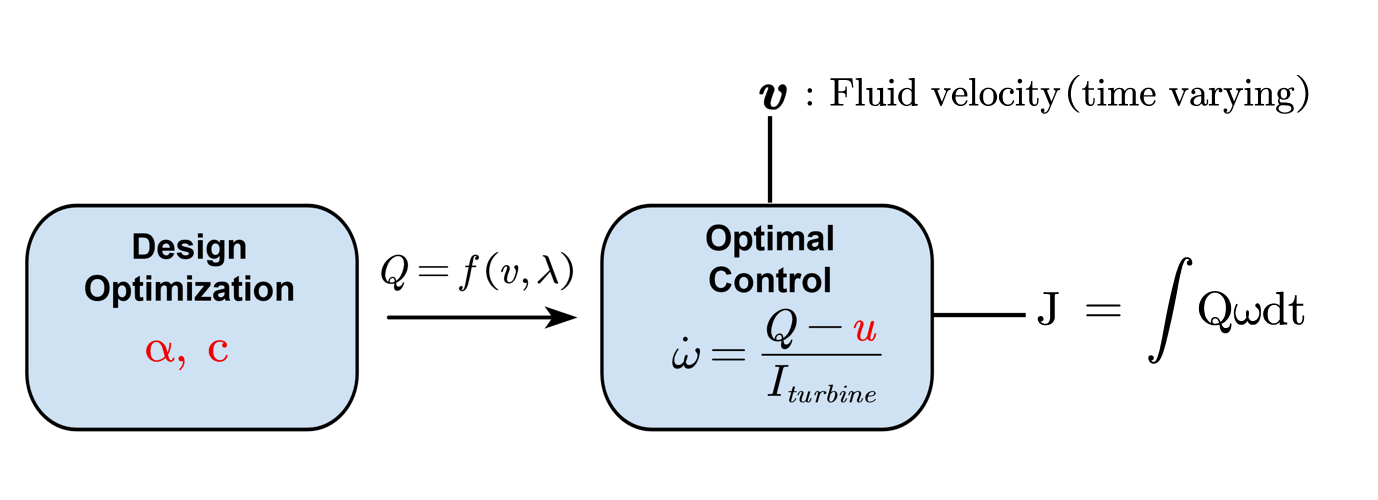}\vspace{-10pt}
      \caption{Illustration of the sequential design-control process.}\vspace{-10pt}
            \label{figure: seq}
\end{figure}

The sequential design process, as shown in Figure~\ref{figure: seq}, consists of two steps: design optimization and optimal control. The design optimization is conducted first by focusing on steady-state power coefficient $C_p$, by optimizing twist angle $\alpha$ and chord length $c$ along the radius. $C_p$ is defined as the ratio between the theoretical output energy of an HKT and the total available flow energy. The design optimization formulation is defined as:
\vspace{-10pt}
\begin{maxi}|l|
    {c_i,\alpha_i}{C_p}{}{}
    \addConstraint{0}{< c_i \leq 1~m}{\quad i = 1,2,..., N}
    \addConstraint{0}{\leq \alpha_i \leq 30^\circ}{\quad i = 1,2,..., N}
     \label{Equ1}
\end{maxi}
where, N is the number of discretized sections on turbine blades. Once \eqref{Equ1} is solved, the optimal control is applied in the second step to maximize the energy generation. Note that the power generated by the rotor is given by 
\vspace{-10pt}
\begin{equation}
    P = Q \omega
\end{equation}
where the flow-induced torque $Q$ is a function of velocity \textit{v} and tip speed ratio $\lambda=\frac{\omega r}{v}$, where \textit{r} is the turbine radius. 
\vspace{-20pt}

For control design, a simple dynamic model, representing turbine kinematics is used, which assumes the drivetrain is rigid and its energy loss is negligible. $Q$ and $\lambda$ are passed from the physical design to the next step of the control design. The rotor rotating speed $\omega$ is determined by the rotor dynamics: 
\vspace{-10pt}
\begin{equation}
    I_{turbine} \dot{\omega}=Q-u
\end{equation}
where \textit{u} is the generator torque that is treated as a control variable in this study. $I_{turbine}$ is the rotor inertia. In this study, the generator dynamics are ignored. The detailed optimal control formulation for the time period $[0,t]$ is shown in (\ref{Equ: optcontrol}):
\vspace{-10pt}
\begin{maxi}
    {u_{[0,t]}}{J= \int_{0} ^t Q \omega dt}{}{}
    \addConstraint{\dot{\omega}}{=\frac{Q-u}{I_{turbine}}}
    \addConstraint{0}{\leq u_{[0,t]} \leq u_{max}}
    \addConstraint{0}{\leq \omega }
    \label{Equ: optcontrol}
\end{maxi}
where $u_{max}$ is the maximum value of the generator torque. The turbine is constrained to have a positive rotating speed, i.e., it rotates in one direction only. 

\vspace{-10pt}
\subsection{Control Co-design Formulation}
\begin{figure}[h!]
\centering
      \includegraphics[scale=0.4]{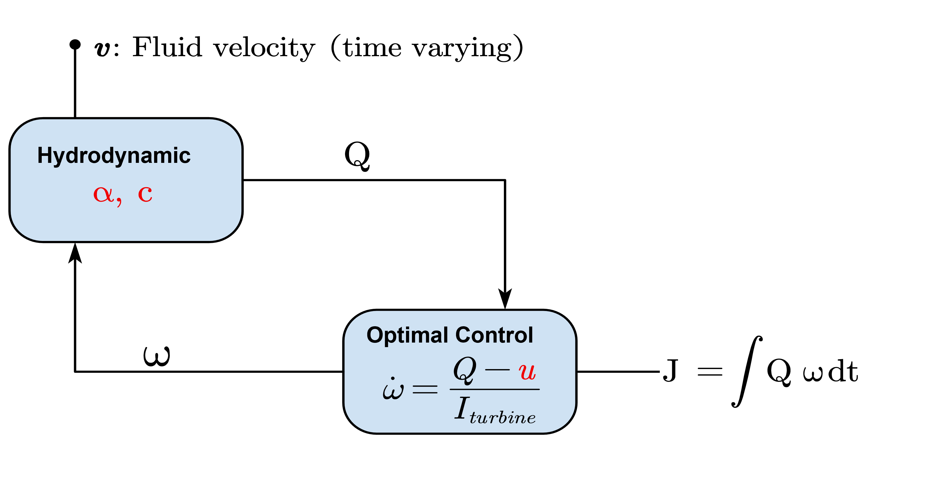}\vspace{-10pt}
      \caption{Illustration of the CCD framework.}\vspace{-10pt}
       \label{fig: CCD}
\end{figure}

CCD provides a framework to simultaneously optimize physical design variables and control parameters for best performance and trade-offs. In this manuscript, the CCD uses a coupled model consisting of hydrodynamic analysis and optimal torque control. The detailed CCD formulation for the time period $[0,t]$ is shown in~\ref{fig: CCD}, note that this formulation is defined for a specific time-varying flow profile over a time period [0,t].
\vspace{-20pt}

The detailed formulation of CCD is given by (\ref{Equ:CCD})
\vspace{-10pt}
\begin{maxi}|l|
    {u_{[0,t]},c_i,\alpha_i}{J= \int_{0} ^t Q \omega dt}{}{}
    \addConstraint{\dot{\omega}}{=\frac{Q-u}{I_{turbine}}}
    \addConstraint{Q}{=g(c_i,\alpha_i, v ,\lambda)}
    \addConstraint{0}{< c_i \leq 1~m}{\quad i = 1,2,..., N}
    \addConstraint{0}{\leq \alpha_i \leq 30^\circ}{\quad i = 1,2,..., N}
    \addConstraint{0}{\leq u_{[0,t]} \leq u_{max}}
    \addConstraint{0}{\leq \omega }
    \label{Equ:CCD}
\end{maxi}
where, the fluid-induced torque $Q$ is a function of $c_i,\alpha_i, v $, and $\lambda$.

\section{COMPARATIVE CASE STUDIES AND RESULTS}

Case studies for CCD with different control constraints are investigated together with sensitivity analysis to flow profiles and initial geometries. Details, including physical designs, power production, trajectories of optimal control, and turbine rotating speed, are reported and discussed in this section. Geometries and power generation of CCD designs are compared against those of sequential designs and the baseline under the same constraints and fluid conditions. We categorize the case studies into two scenarios. 
In the first scenario, we assume there is no upper limits on the control, i.e., $u_{max} = \infty$. The unlimited control load assumption is impractical, given the limitations on the size and rated power of the generator. The case studies are conducted to identify and understand the couplings between the physical design space and the control design space.
In the second scenario, the upper limit of control load is set as the generator static recorded in~\cite{jonkman2009definition}, that is, $u_{max}=47000$ Nm.

\vspace{-10pt}
\subsection{Scenario 1: Without constraint on control load}
 Comparative HKT design optimization case studies are conducted using CCD and a sequential approach without constraint on control load. The baseline and initial geometry for both design optimization methods are defined as the scaled HKT model introduced in Section 2.1. The cases are first investigated with a sinusoidal flow profile, which is defined as $v=1.4+0.2 \sin (0.1t)$, for a time span of 150 seconds. 

\begin{figure}[h!]
      \includegraphics[scale=0.6]{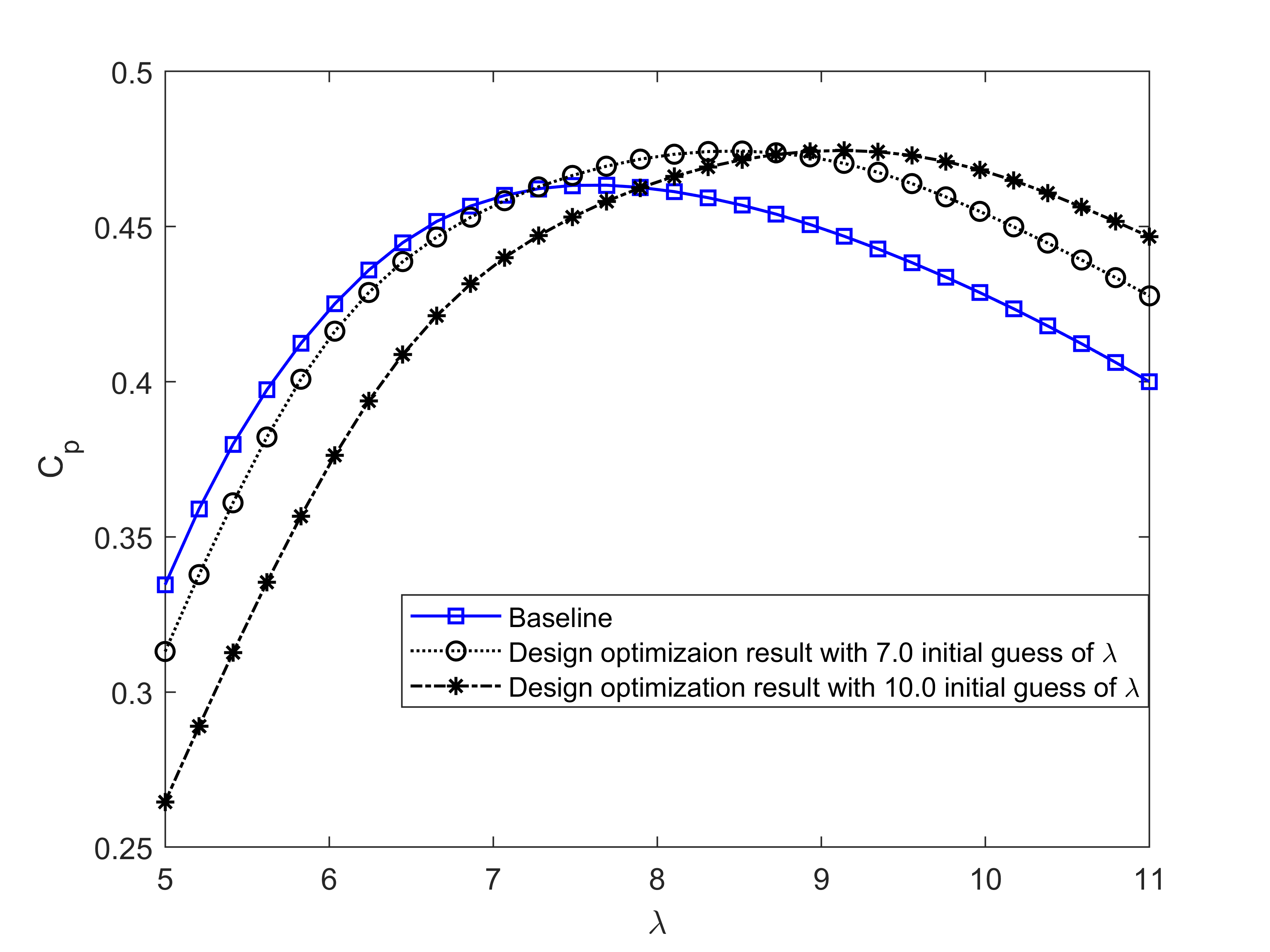}\vspace{-10pt}
      \caption{Comparisons of power coefficients between the baseline geometry and the sequential design.}\vspace{-10pt}
       \label{fig: Cp}
\end{figure}

The steady-state power coefficient $C_p$ of the baseline geometry and the design optimization geometries are reported in Figure~\ref{fig: Cp}. The baseline geometry has the maximum $C_p$ of 0.4633 at $\lambda=7.6$. The design optimization in the sequential process yields two different geometries according to the $\lambda$ initial guessed values, with the first one having a maximum power coefficient of 0.4744 at $\lambda=8.5$; and the second one having a slightly higher maximum power coefficient of 0.4745 at $\lambda=9.1$.
%

\begin{figure}[h!]
      \includegraphics[scale=0.6]{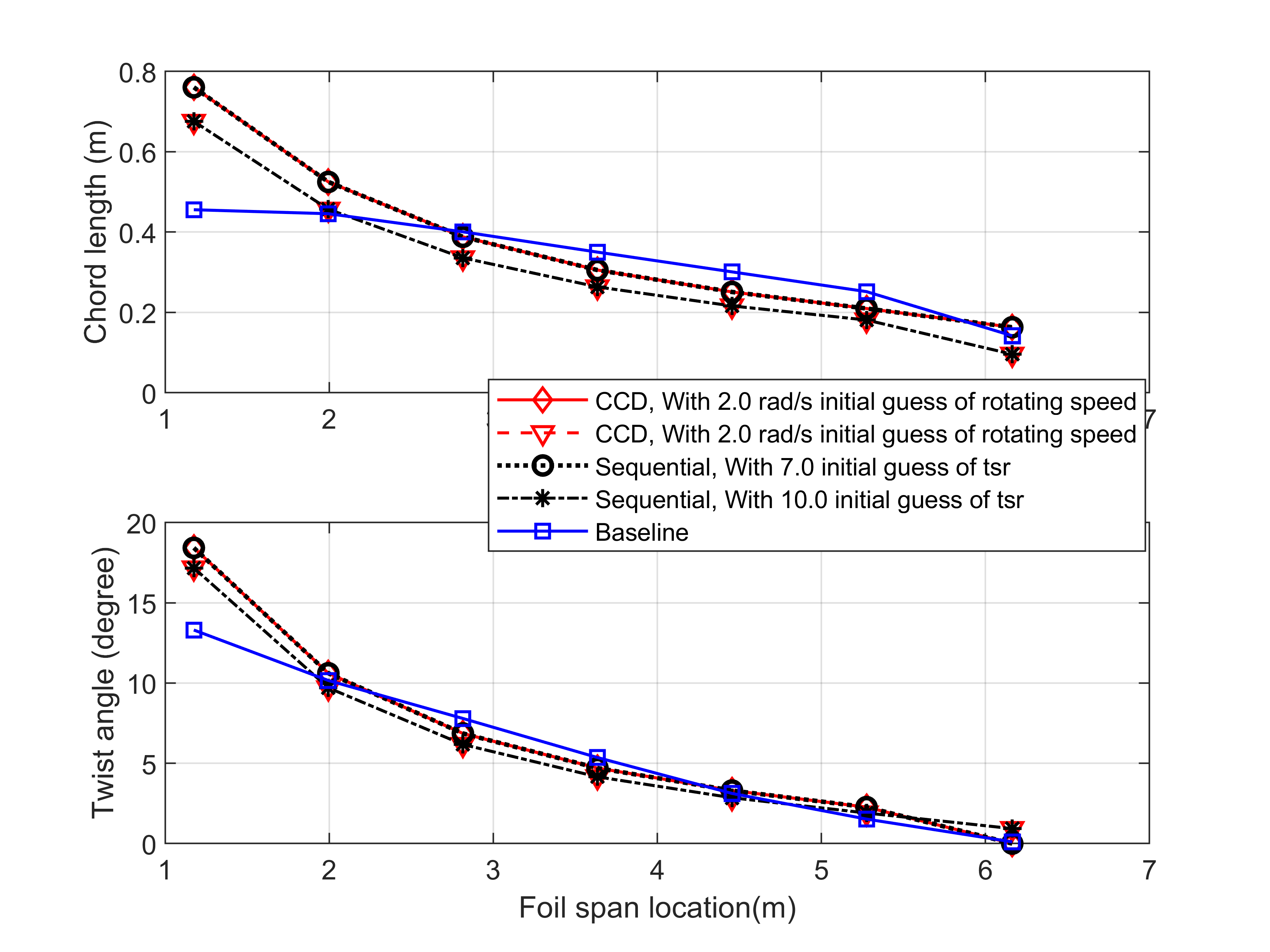}\vspace{-10pt}
      \caption{Comparison of physical geometry optimized with CCD, sequential design, and baseline geometry without constraint on control load.}\vspace{-14pt}
       \label{figure04}
\end{figure}

Since we observed two different design optimization results when starting from different $\lambda$, it is possible that control co-design could also be sensitive to turbine rotating speed initial guess. 
Therefore, CCD with different turbine rotating speed initial guesses are tested, and the optimized physical geometries and the distributions of chord lengths and twist angles are compared to those of the sequential designs and the baseline. 
As shown in Figure~\ref{figure04}, the CCD shows sensitivities to the initial guess of rotating speed.
With different initial guesses of rotational speed, the final CCD results converge to different results. 
Furthermore, the CCD design and design from sequential design are identical, respectively, for both cases. 
The identical results demonstrate that, in the absence of coupling between the design and control spaces, the sequential design problem and CCD lead to similar physical design.
%

\begin{figure}[h!]
      \includegraphics[scale=0.6]{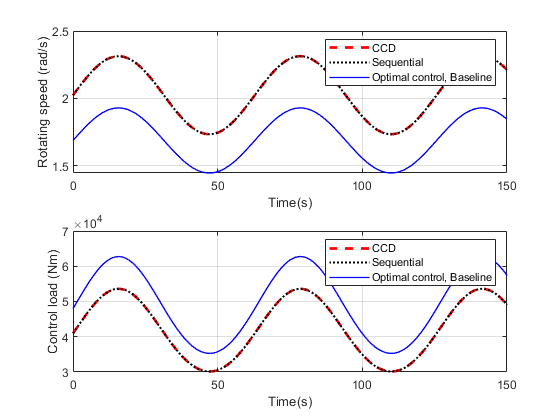}\vspace{-10pt}
      \caption{Comparison of turbine rotating speed and control loads for CCD, sequential, and baseline without constraint on control load for sinusoidal inflow profile.}\vspace{-10pt}
      \label{figure3}
\end{figure}

Since there are two optimized designs, we choose the better one for both approaches for a more detailed discussion.
The dynamical responses of the turbines designed using sequential and CCD approaches are shown in Figure~\ref{figure3}.
Without constraint on the amplitude of control load, the optimal control can find the control trajectories to maintain the optimal turbine tip speed ratio, thereby decoupling the physical design optimization process from the control optimizations.
In addition to the physical design, the control trajectories of the optimized designs are also identical between the two design approaches.
The key parameters, including computation time and generated energy, of both design methods, are shown in Table~\ref{table_2} and compared to that of the baseline design. An improvement of 2.4\% in energy production compared to the baseline is obtained. It should be noted that as CCD expands the search space in the optimization process, its computational demand is substantially higher compared to the sequential design process. The calculations are conducted using ~3.4G Hz AMD Ryzen 5950X. 
\vspace{-20pt}

\begin{table}[h!]
\caption{Comparisons on energy production between (1) baseline geometry with optimal control, (2) sequential design with optimal control, and (3) CCD without constraint on control load.}\vspace{-6pt}
\label{table_2}
\begin{center}
\begin{tabular}{l c c c }
\hline
\textbf{Design} & \textbf{Computation} & \textbf{Generated}  & \textbf{Improvement} \\
\textbf{Method}  & \textbf{Time ($s$)} & \textbf{Energy} ($kJ$) & (\%) \\
\hline 
Baseline & 1373 & 13165 & 0 \\
Sequential  & 1495 & 13481 & \textcolor{blue}{+2.4} \\
CCD & 2469 & 13481 &  \textcolor{blue}{+2.4} \\
\hline
\end{tabular}\vspace{-16pt}
\end{center}
\end{table}

\begin{figure}[h!]
      \includegraphics[scale=0.6]{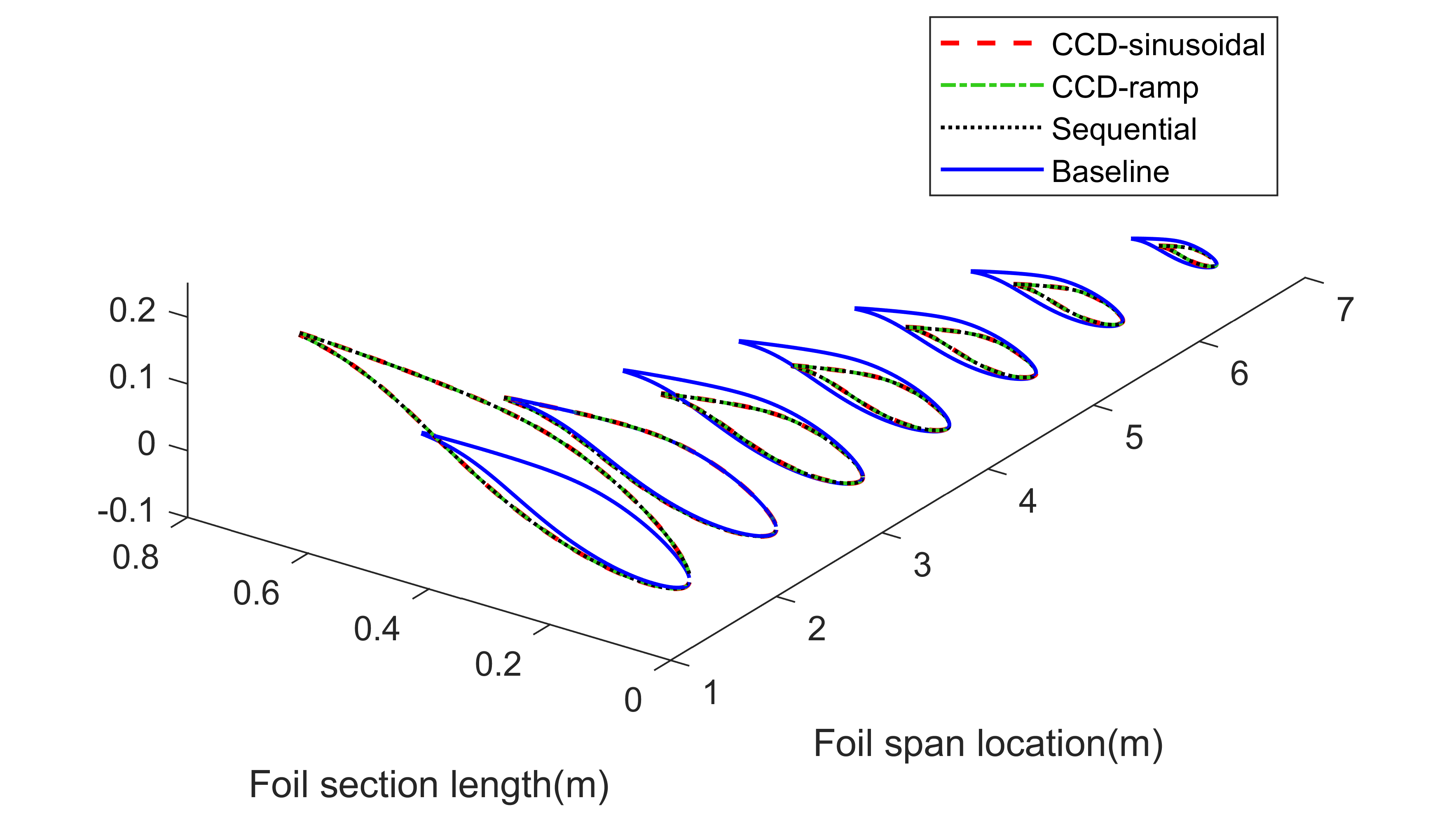}\vspace{-10pt}
      \caption{Comparison of physical geometry optimized based on
different flow profiles without the constraint on control load.}\vspace{-12pt}
      \label{figure5}
\end{figure}

\begin{figure}[h!]
      \includegraphics[scale=0.6]{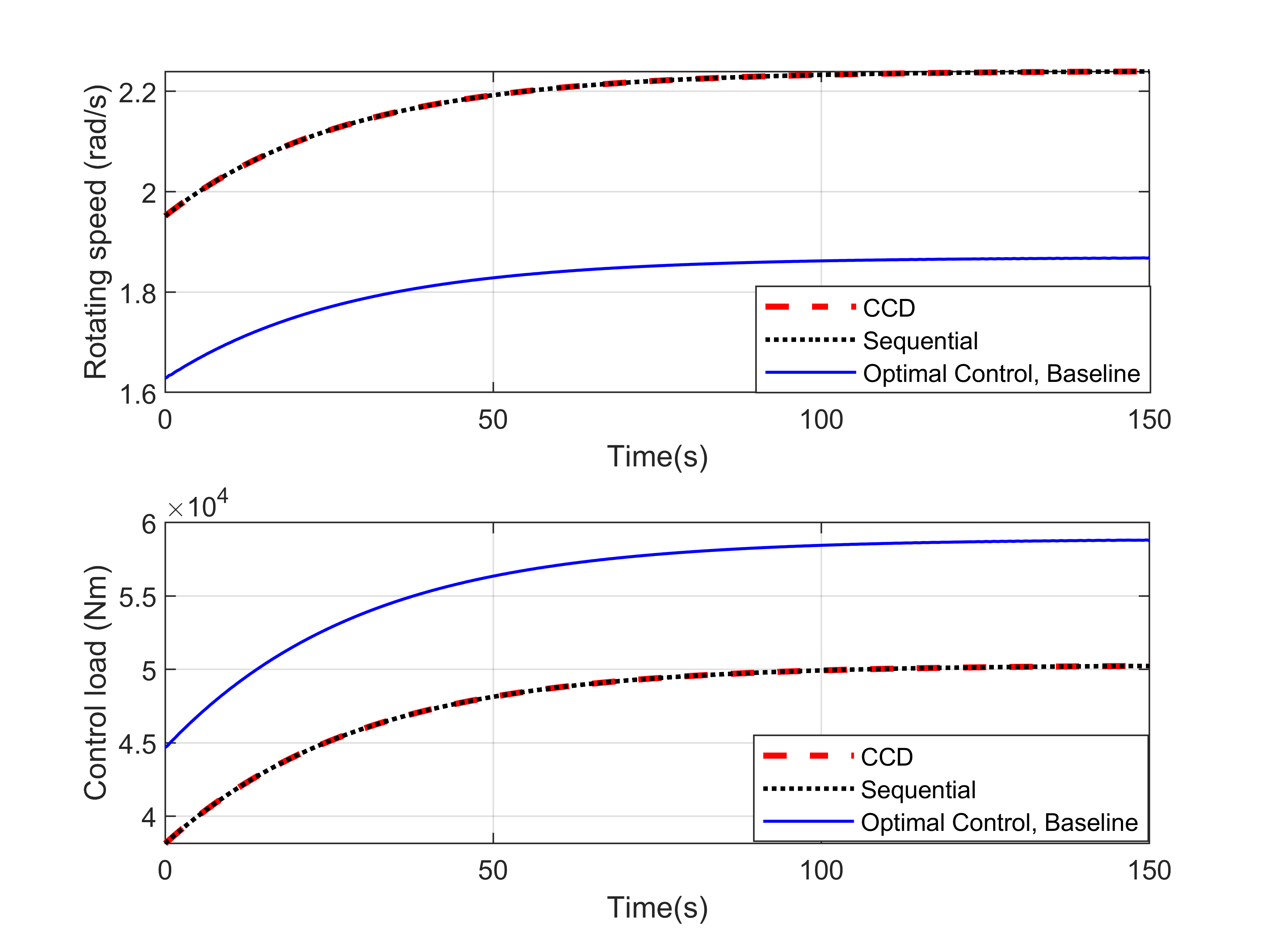}\vspace{-10pt}
      \caption{Comparison of turbine rotating speed and control loads for CCD, sequential, and baseline without constraints on control loads for a ramp flow profile.}\vspace{-12pt}
      \label{figure6}
\end{figure}

To investigate if CCD and sequential designs are sensitive to the flow profile, we perform additional optimization with a different ramp flow profile: $v=1.55-0.2 (0.1t)^{0.7}$. The final CCD design with this different flow profile converges to the same geometry, as shown in Figure~\ref{figure5}, and the CCD and the sequential methodology yield the same control trajectories as shown in Figure~\ref{figure6}.  Without any control load constraints, the controller can ensure that optimal $\lambda$ is achieved, regardless of the flow speed profile. 


\begin{figure}[h!]
      \includegraphics[scale=0.6]{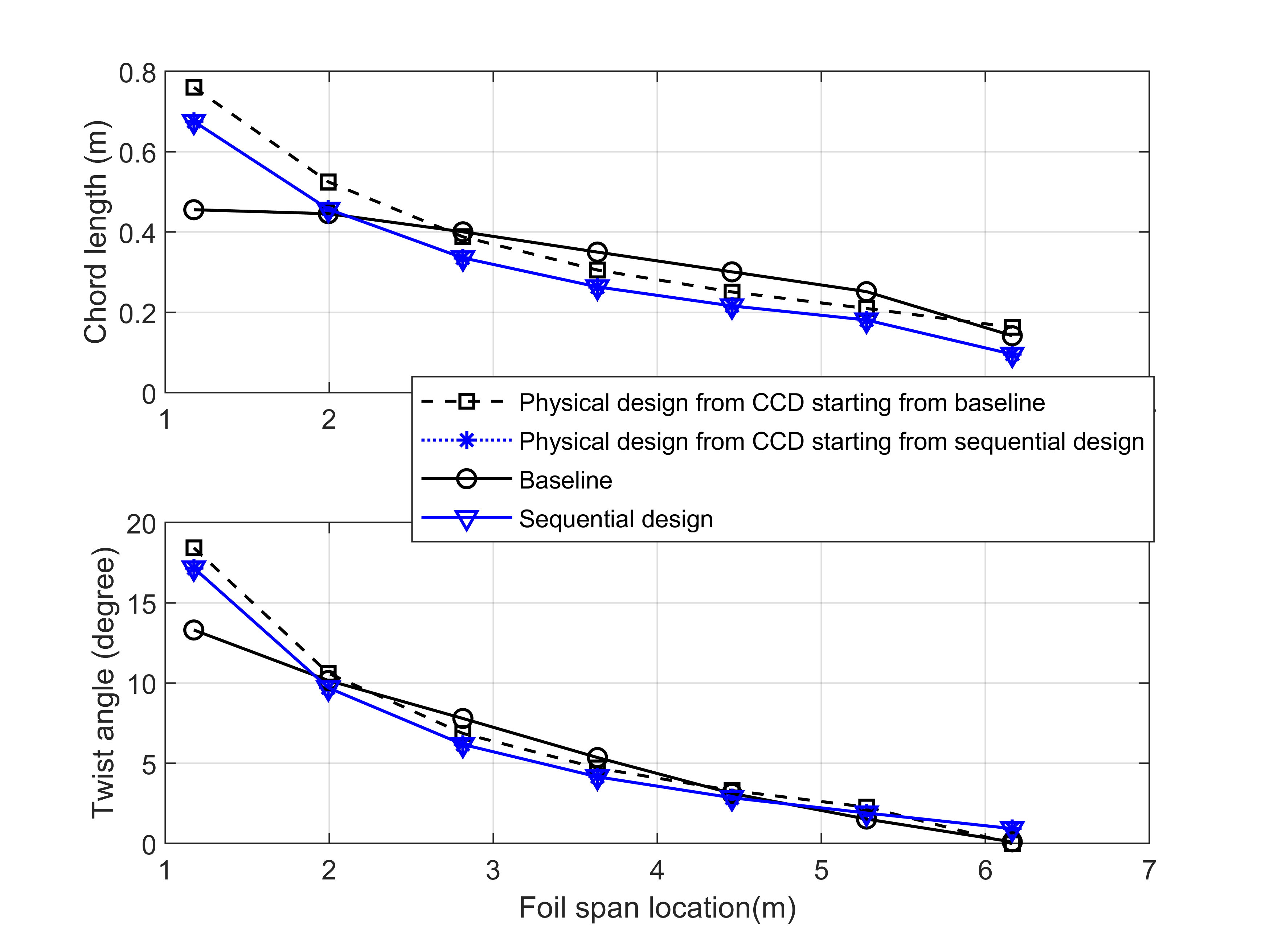}\vspace{-10pt}
      \caption{Comparison of physical geometry with different starting geometry.}\vspace{-12pt}
      \label{figure: sensitivity1}
\end{figure}

The sensitivity of CCD to the initial geometry is further investigated by comparing the physical designs from CCD with different initial designs, i.e., the baseline and sequential design. Case studies using the flow profile: $v=1.4+0.2 \sin (0.1t)$. Different initial physical design parameter values and respective optimized values are shown in Figure~\ref{figure: sensitivity1}. According to the case studied, the physical design from the current CCD problem are both sensitive to the initial condition and are identical to those from the sequential designs reported in Figure~\ref{fig: Cp}.
%

\subsection{Scenario 2: With constraint on control load}
\begin{figure}[h!]
      \includegraphics[scale=0.6]{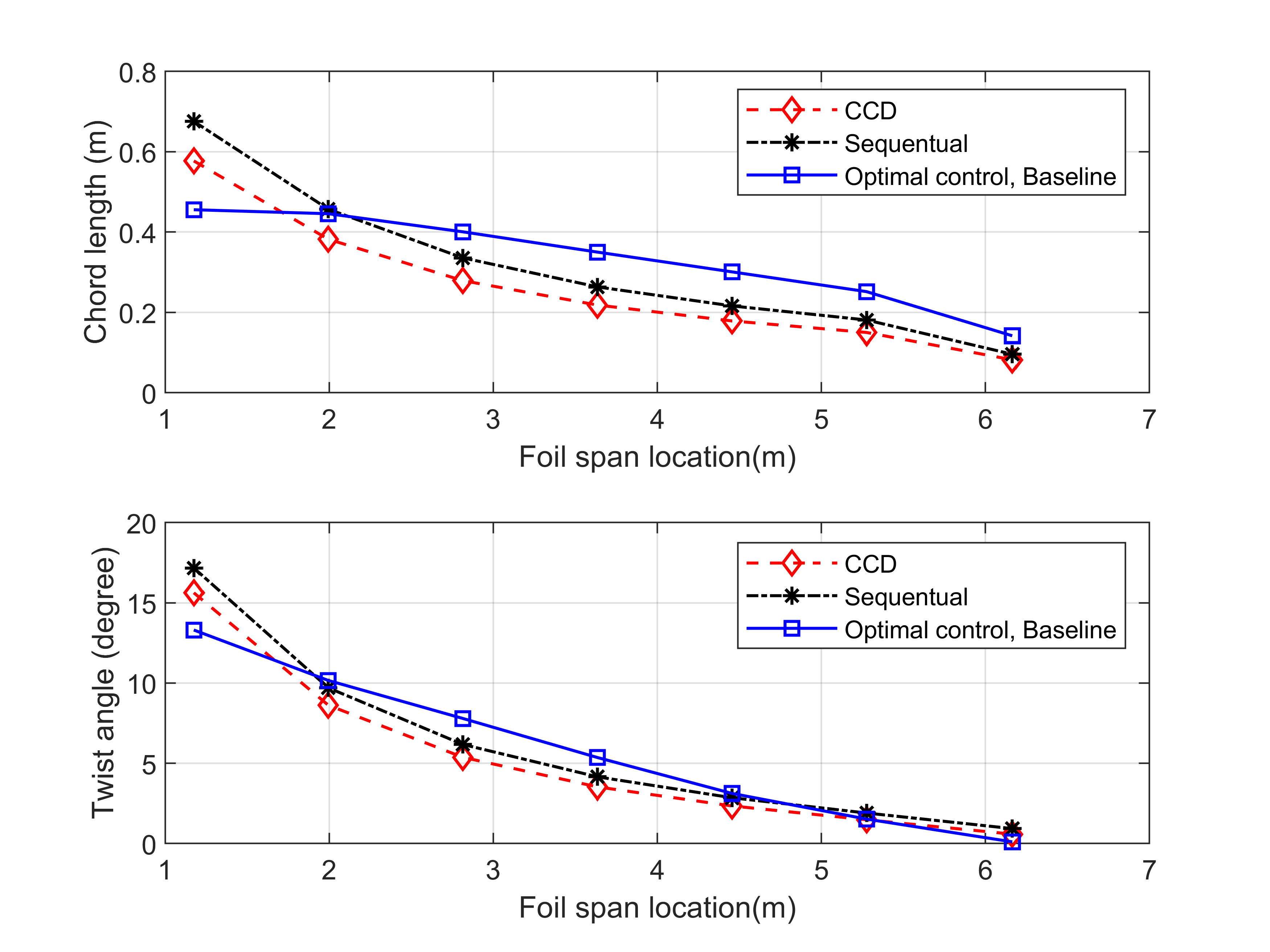}\vspace{-10pt}
      \caption{Comparison of physical geometry optimized with CCD, sequential design, and baseline geometry with constraint on control load.}\vspace{-12pt}
      \label{figure:geometry with constraints}
\end{figure}

Comparative HKT design optimization case studies are conducted using CCD and a sequential approach with a constraint on control load when the control \textit{u} is constrained based on the data of generator reported by Jonkman et al.~\cite{jonkman2009definition}. The cases are investigated with the same sinusoidal flow profiles as in scenario 1 described in 4.1. The initial geometry for both methods is the scaled HKT model introduced in Section 2.1.
The optimized physical geometries and the distributions of chord lengths and twist angles are shown in Figure~\ref{figure:geometry with constraints}. CCD does not yield the same physical geometry as that in the sequential method. Instead, CCD converges to a physical geometry with a relatively shorter chord length and smaller twist angle, achieving a maximum power coefficient of 0.4745 at $\lambda=10.0$. Since the design optimization in the sequential process does not take control constraints into consideration, the physical design is the same as in scenario 1. But the optimal control is different from that shown in Figure~\ref{figure3}.
\vspace{-20pt}

\begin{figure}[h!]
      \includegraphics[scale=0.6]{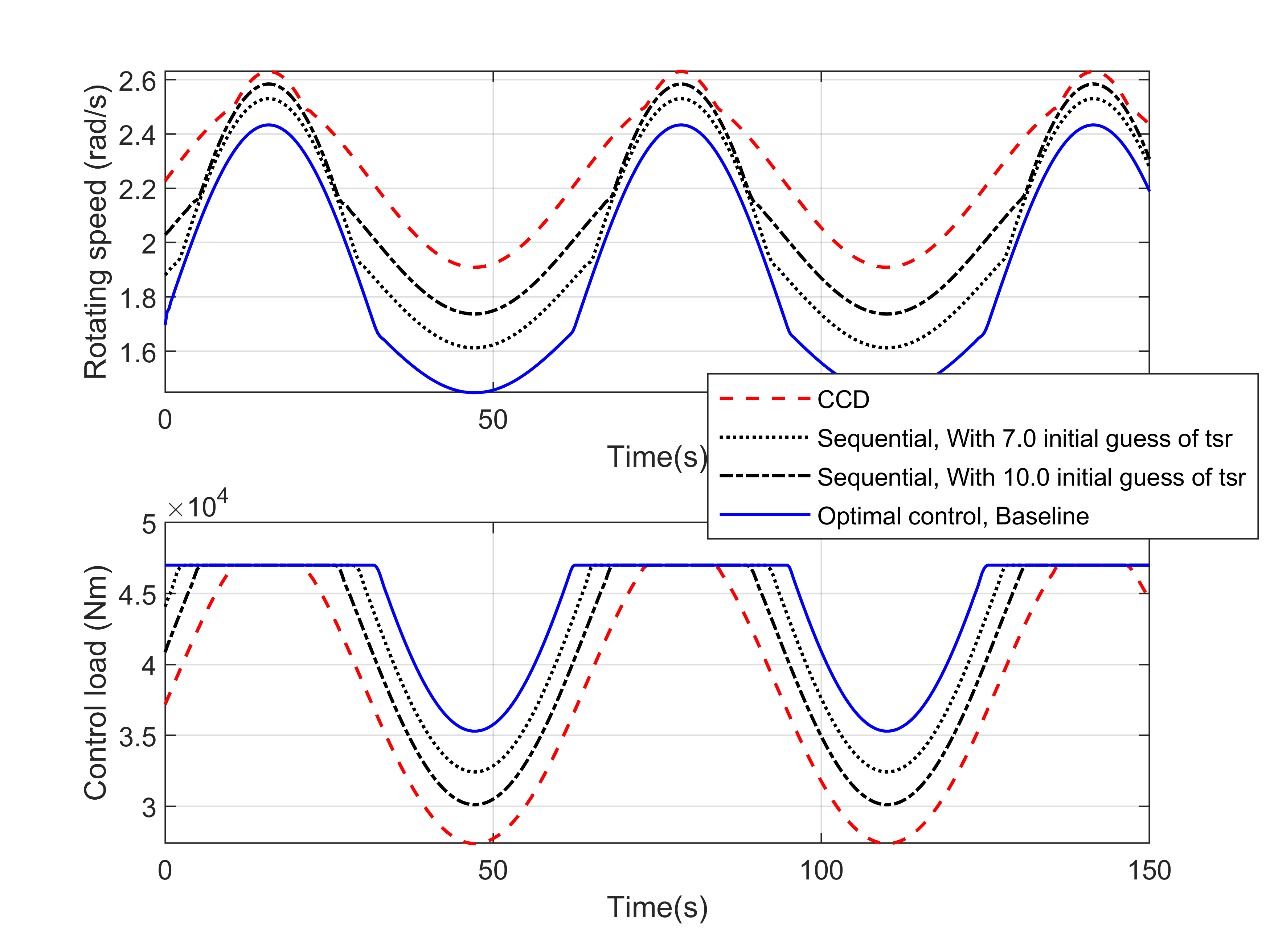}\vspace{-10pt}
      \caption{Comparison of turbine rotating speed and control loads for CCD, sequential, and baseline with constraint on control load for a sinusoidal flow profile}\vspace{-20pt}
      \label{figure:history with const}
\end{figure}

The turbine rotating speeds and control responses of different designs are shown in Figure~\ref{figure:history with const}. 
 Since CCD optimizes the physical geometry parameters and control load trajectory simultaneously, the physical design will be affected by the control load constraint. As a consequence, the CCD design rotates at a higher speed compared to that of a sequential design and reduces the time for the system to operate in control load saturation conditions.
Even though the CCD design has a similar maximum power coefficient as compared to that of the sequential design, the overall output energy, as shown in Table~\ref{table_3}, is 4.9\% higher than the baseline versus 4.3\% improvement achieved by the sequential design. 

The influence of the control load constraints on the optimized physical geometry shows that the control constraint leads to a coupling between control and design spaces.
With this coupling, the CCD has the advantage of finding the optimal design that might not be reached with the sequential design approach.
\vspace{-20pt}
\begin{table}[h!]
\caption{Comparisons on energy production between (1) baseline geometry with optimal control, (2) sequential design with optimal control and, (3) CCD with constraint on control load.}\vspace{-5pt}
\label{table_3}
\begin{center}
\begin{tabular}{l c c c }
\hline
\textbf{Design} & \textbf{Computation} & \textbf{Generated}  & \textbf{Improvement} \\
\textbf{Method}  & \textbf{Time ($s$)} & \textbf{Energy} ($kJ$) & (\%) \\
\hline 
Baseline & 288 & 12839 & 0 \\
Sequential & 355 & 13398 & \textcolor{blue}{+4.3} \\
CCD & 1470 & 13464 &  \textcolor{blue}{+4.9} \\
\hline
\end{tabular}\vspace{-12pt}
\end{center}
\end{table}

\begin{figure}[h!]
      \includegraphics[scale=0.6]{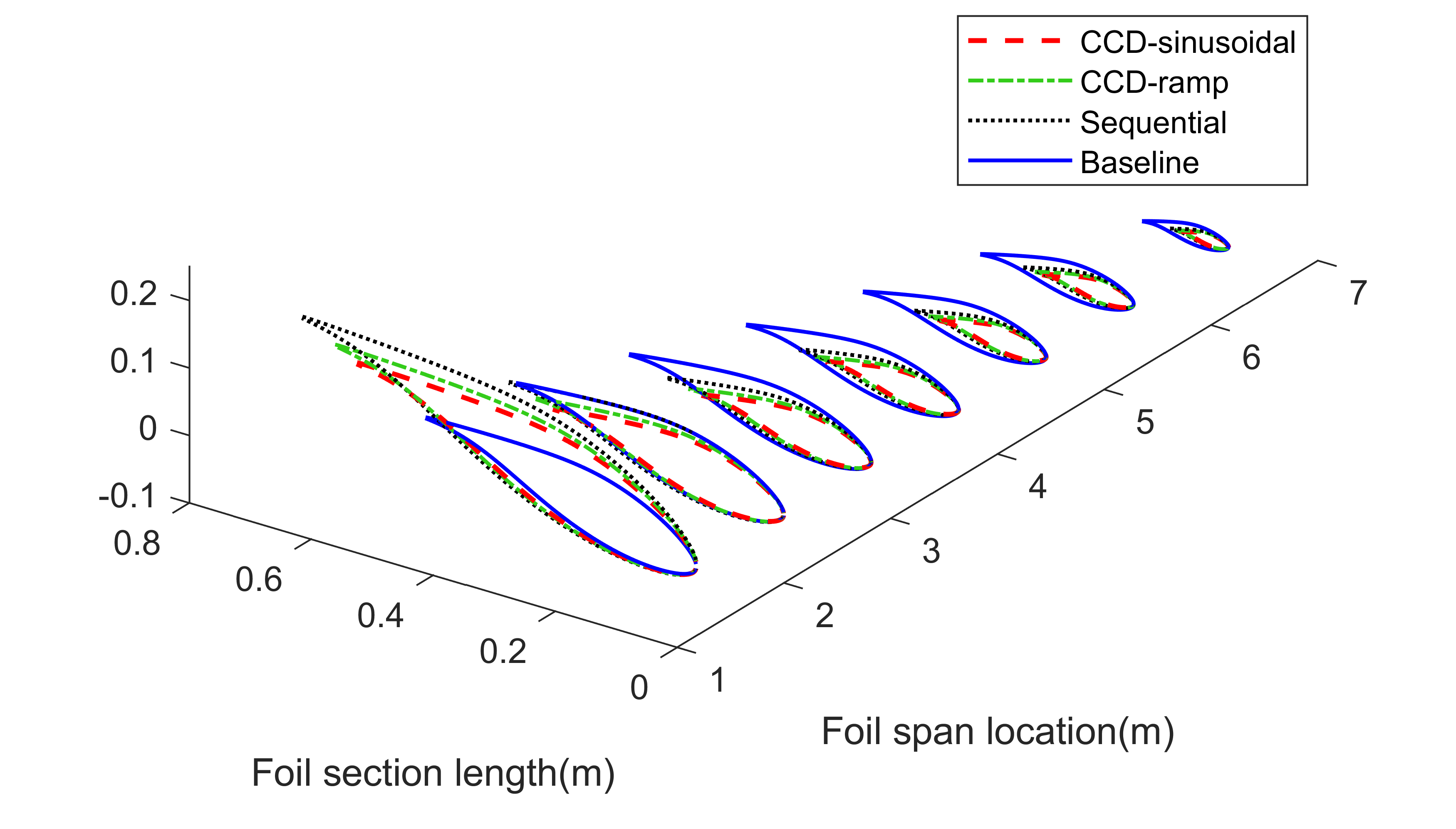}\vspace{-10pt}
      \caption{Comparison of physical geometry optimized based on different flow profile with control load.}
      \label{figure: sensitivity with cons}\vspace{-10pt}
\end{figure}

The sensitivity of CCD to flow profiles is investigated with a ramp flow profile: $v=1.55-0.2 (0.1t)^{0.7}$. As shown in Figure~\ref{figure: sensitivity with cons}, the CCD designs converge to different geometries with different flow profiles. As the control actions are different for different flow profiles, the coupling between the control design and physical spaces necessitates the adjustment in the design space to maintain the optimal outcome. This observation, however, leads to a challenge in CCD: How to select the fluid profiles so that CCD will yield a single optimal design under the most possible operating scenario? This will be a topic of our future work.
%
\begin{figure}[h!]
      \includegraphics[scale=0.6]{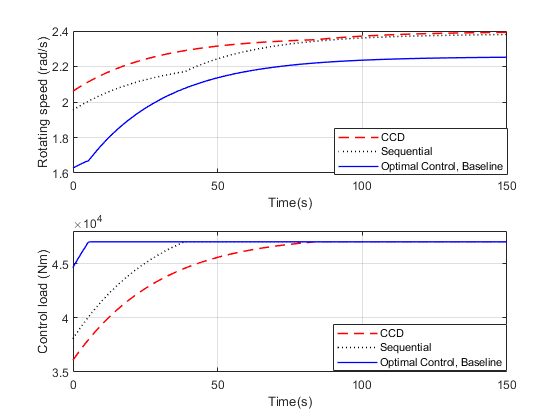}\vspace{-10pt}
      \caption{Comparison of turbine rotating speed and control loads for CCD, sequential, and baseline with constraint on control load for a ramp velocity profile.}\vspace{-10pt}
      \label{figurelabel}
\end{figure}

\begin{figure}[h!]
\centering
      \includegraphics[scale=0.7]{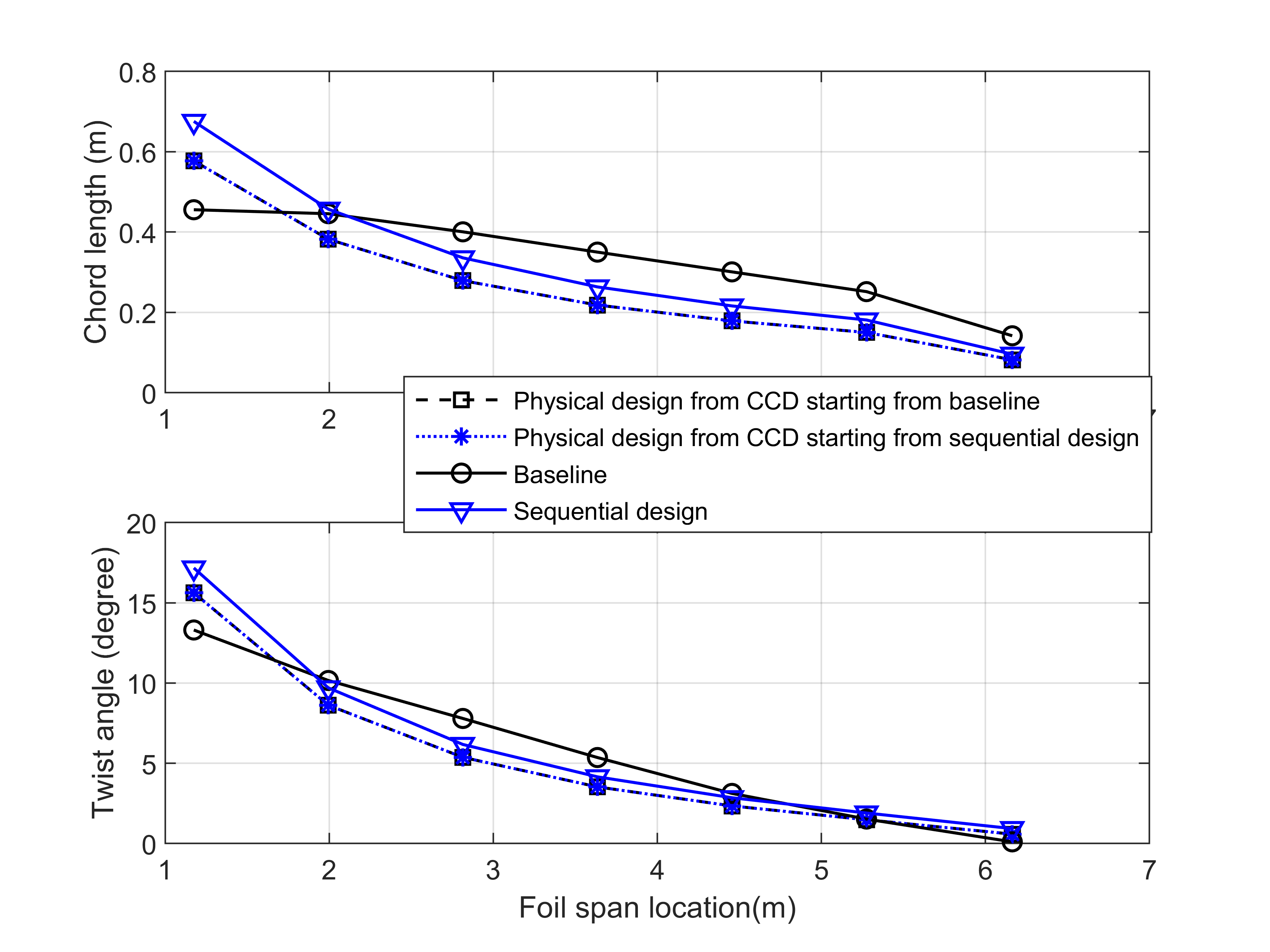}\vspace{-10pt}
      \caption{Comparison of physical geometry with different starting geometry considering constraint on load.}\vspace{-10pt}
      \label{figure: sensitivity with constraints}
\end{figure}

We perform comparative case studies similar to that described in scenario 1 to investigate the sensitivity of CCD to the initial geometry, but with a constraint on control load. Physical design parameters with different initial values and respective optimized values are shown in Figure~\ref{figure: sensitivity with constraints}. With different initial geometries, the final CCD designs are identical within the tolerance.

\vspace{-8pt}
\section*{CONCLUSION}
This paper introduced a CCD framework for the optimization of the rotor of an HKT. A coupled dynamic-hydrodynamic model was introduced, based on which the CCD problem was formulated to maximize the energy production with time-variant flow profiles. A comprehensive comparison between the CCD and the sequential design was carried out to reveal the interaction between the physical design space and the control design space. The comparison study also provides insights for system integration design. The optimized geometries and energy production result shows that: in the absence of control constraints, no coupling between the physical design space and the control space was identified. The sequential design method and CCD lead to identical design and power output. With control constraints included in the CCD, the coupling between design and control spaces exists and CCD yields better performance compared to the sequential approach. The numerical efficiency and robustness of the CCD tools were also evaluated. Our optimization studies show that for a complex system like HKT, control co-design is essential to yielding the optimal design and performance.

\vspace{-8pt}
\section*{ACKNOWLEDGEMENT}
This work is supported by the United States
Department of Energy (DOE)-ARPA-E under SHARKS program award No. DE-AR0001438.

\vspace{-8pt}
\bibliographystyle{asmems4}
\bibliography{asme2e}

\end{document}